%% file: note2421.tex
\documentclass[twocolumn,showpacs,aps,prl,superscriptaddress,preprintnumbers]{revtex4}

\long\def\inst#1{\par\nobreak\kern 4pt\nobreak
    {\it #1}\par\vskip 10pt plus 3pt minus 3pt}

\usepackage[dvips]{graphicx}
\usepackage{epsf,epsfig,rotating,color,colordvi,relsize,xspace}

\RequirePackage{xspace}

\input plb_symbols
\input mysymbols

\begin{document}

%
%
\preprint{\babar-PUB-11/026}
\preprint{SLAC-PUB-14868}
\preprint{arXiv:1202.3650}

\title{
\large \bfseries \boldmath Search for lepton-number
violating processes in $\Bp \to h^- \ellp\ellp$ decays
}

\input authors_dec2011


\begin{abstract}
We have searched for the lepton-number violating processes $\Bp\to h^-
\ellp\ellp$ with $h^-=\Km/\pim$ and $\ellp=e^+/\mup$, using a sample
of \nbb\ million \BB\ events collected with the \babar\ detector at
the \pep2\ $\epem$ collider at the SLAC National Accelerator
Laboratory. We find no evidence for these decays and place 90\%
confidence level upper limits on their branching fractions $\calB\
(B^+\rightarrow\pi^- e^+e^+) < 2.3 \times 10^{-8}$, $\calB\
(B^+\rightarrow K^- e^+e^+) < 3.0 \times 10^{-8}$, $\calB\
(B^+\rightarrow\pi^- \mu^+\mu^+) < 10.7 \times 10^{-8}$, and $\calB\
(B^+\rightarrow K^- \mu^+\mu^+) < 6.7 \times 10^{-8}$.
\end{abstract}


\maketitle


In the Standard Model (SM), lepton number $L$ is conserved in
low-energy collisions and decays~\cite{bib:sphaleron} and the lepton
flavor numbers for the three lepton families are conserved if
neutrinos are massless. The observation of neutrino
oscillations~\cite{bib:neutrinos} indicates that neutrinos have mass. If the
neutrinos are of the Majorana type~\cite{bib:majorana}, the neutrino
and antineutrino are the same particle and processes that involve
lepton-number violation become possible. The lepton number must change
by two units ($\Delta L=2$) in this case and the most sensitive
searches have so far involved neutrinoless nuclear double beta decays 
$0\nu\beta\beta$~\cite{bib:bb}. The nuclear environment complicates
the extraction of the neutrino mass scale. Processes involving
meson decays have been proposed as an alternative that can also look
for lepton-number violation with muons or $\tau$ leptons.

\begin{figure}[ht!] 
\begin{center} 
\begin{tabular}{c}
   \epsfig{file=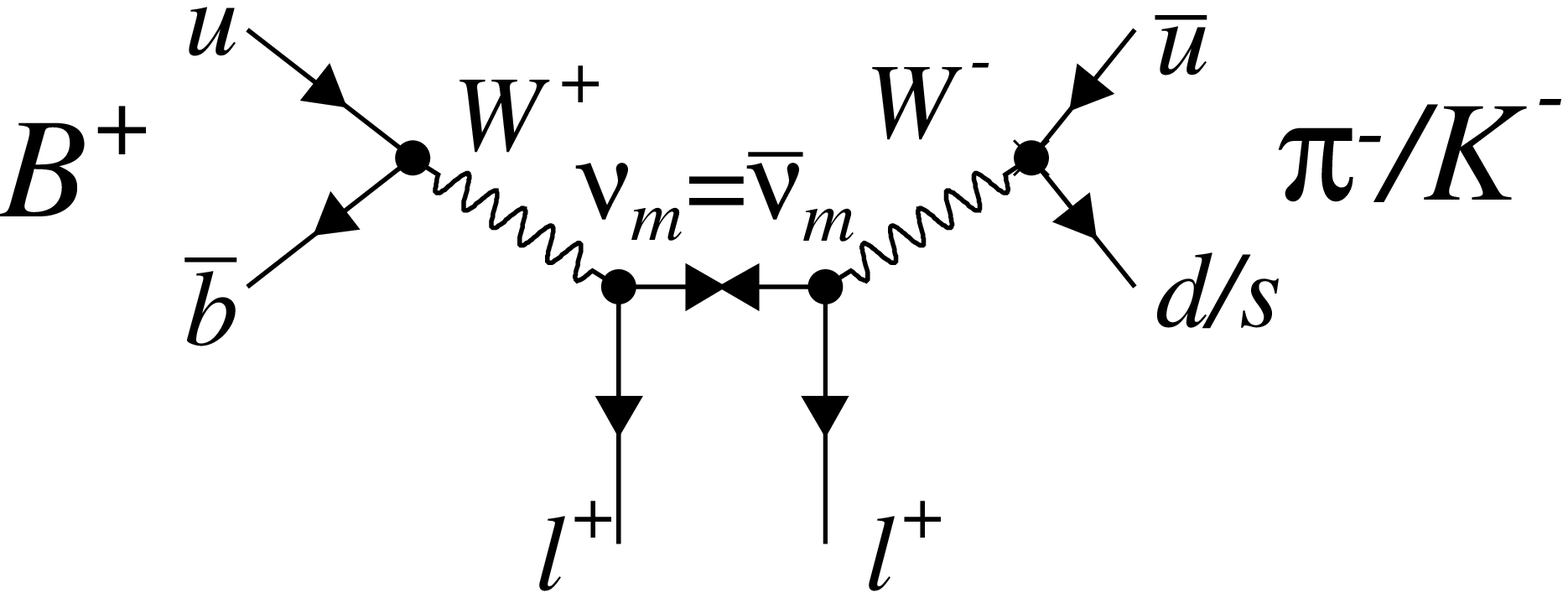,width=0.8\columnwidth} 
\end{tabular}
\caption{An example diagram of the $\Delta
  L=2$ process $\Bp\to h^-\ellp\ellp$ via $s$-channel Majorana neutrino $\nu_{m}$
  production and decay.}
\label{fig:feyn}
\end{center} 
\end{figure}

An example of a decay involving mesons is $\Bp\to h^-\ellp\ellp$,
where \ellp = \ep or \mup\ and $h^-$ is a meson with a mass smaller
than the \Bmeson.  A possible mechanism for this process involving the
production and subsequent decay of a Majorana neutrino $\nu_m$ is
illustrated in Fig.~\ref{fig:feyn}, which is topologically similar to
the $t$-channel Feynman diagram in $0\nu\beta\beta$ decays. If the Majorana
neutrino mass lies between the $h$ meson and the \Bmeson\ masses,
resonance production could result in an enhanced peak in the invariant
mass spectrum of the hadron and one of the leptons~\cite{bib:atre}.

The experimental approach in searches for these lepton-number
violating processes is very similar to the approach for $\B\to
K^{(*)}\ellp\ellm$ and we use many of the techniques reported in
Refs.~\cite{searches,newbabar} to search for the four modes $\Bp\to
h^- \mup\mup$ and $\Bp\to h^- e^+e^+$, where $h^-=\Km$ or \pim
(charge-conjugate modes are implied throughout this paper). Previous
searches for these decays have produced 90\% confidence level (C.L.)
upper limits on the branching fractions in the range $(1.0-8.3) \times
10^{-6}$~\cite{bib:cleo}. The LHCb collaboration recently reported 95\%
C.L. upper limits on the branching fractions $\calB(\BtoKmm) < 5.4
\times 10^{-8}$ and $\calB(\BtoPimm) < 1.3 \times
10^{-8}$~\cite{bib:lhcb}. The Belle collaboration places 90\%
C.L. upper limits on the branching fractions
$\calB(\Bp\to\Dm\ellp\ellp)$ in the range $(1.1-2.6)\times
10^{-6}$~\cite{bib:belle}.

Our search uses a data sample of \nbb\ million \BB\ pairs collected at
the $\FourS$ resonance with the \babar\ detector 
at the \pep2\ asymmetric-energy $\epem$ collider at the SLAC National
Accelerator Laboratory. The \epem\ center-of-mass (CM) energy is
$\sqrt{s} = 10.58$\gev, corresponding to the mass of the
\FourS\-resonance (on-resonance data). In addition, \offreslumi\ of
data collected 40~\mev\ below the \FourS\-resonance (off-resonance
data) are used for background studies.  We assume equal production
rates of \BpBm\ and \BzBzb\ mesons~\cite{bib:pdg}. The \babar\
detector is described in detail in Ref.~\cite{BaBarDetector}.

Monte Carlo (MC) simulation is used to evaluate the background
contamination and selection efficiencies. The simulated backgrounds
are also used to cross-check the selection procedure and for studies
of systematic effects. The signal channels are simulated by the
EvtGen~\cite{bib:evtgen} package using a three-body phase space
model. We also generate light quark \qqbar\ continuum events
($\epem\to\qqbar$, $q=u,d,s,c$), di-muon, Bhabha elastic
\epem\ scattering, \BB\ background and two-photon
events~\cite{bib:twophoton}.  Final-state radiation is provided by
\photos~\cite{bib:photos}. The detector response is simulated with
\geantfour~\cite{bib:geant4}, and all simulated events are
reconstructed in the same manner as data.

We select events that have at least four charged tracks, the ratio of
the second to zeroth Fox-Wolfram moments~\cite{bib:fox} less than 0.5,
and two same-sign charged leptons each with momentum greater than
$0.3$\gevc\ in the laboratory frame. The total transverse vector
momentum of an event calculated in the laboratory frame must be less
than 4\gevc; the distribution of this quantity peaks at 0.2\gevc\ for
signal events. The two leptons are constrained to come from a single
vertex and an invariant mass $m_{\ellp\ellp} < 5.0$\gevcc is required,
to maintain compatibility with Ref.~\cite{searches}. Electrons and
positrons from photon conversions are removed, where photon conversion
is indicated by electron-positron pairs with an invariant mass less
than 0.03\gevcc\ and a production vertex more than 2\cm\ from the beam
axis.

The charged pions and kaons are identified by measurements of their
energy loss in the tracking detectors, the number of photons recorded
by the ring-imaging Cherenkov detector and the corresponding Cherenkov
angle. These measurements are combined with information from the
electromagnetic calorimeter and the instrumented magnetic-flux return
detector to identify electrons and muons~\cite{BaBarDetector}.

The four-momenta of the electrons and positrons are corrected for
Bremsstrahlung emission by searching for compatible photons. Using
measurements made in the laboratory frame, the photon and electron 
four-momenta are combined if the photon energy $E_{\gamma}$ is greater
than 0.05\gev, the shape of the energy deposit in the electromagnetic
calorimeter is compatible with a photon shower, and the difference in polar
angle between the photon and electron, measured at the point of
closest approach to the beam spot, is less than 0.035 rad. In
addition, the azimuthal angles $\phi$ of the photon $\phi_{\gamma}$,
the lepton $\phi_{\ell}$, and the calorimeter deposit associated with
the lepton $\phi_{c}$, all measured at the primary vertex, must be
compatible with $\phi_{\ell}-0.05<\phi_{\gamma}<\phi_{c}$ for
electrons and $\phi_{c}<\phi_{\gamma}<\phi_{\ell}+0.05$ for positrons.

The two leptons and the hadron track are combined to form a \B
candidate. The \B candidate is rejected if the invariant mass of the
two leptons is in the range $2.85<m_{\ellp\ellp}<3.15$\gevcc\ or
$3.59<m_{\ellp\ellp}<3.77$\gevcc. Although a peaking background in the
$\jpsi$ or $\psi(2S)$ mass regions is not expected, these criteria
maintain consistency with Ref.~\cite{searches}. For the mode \BtoPimm,
the invariant mass of each muon and the hadron must be outside the
region $3.05<m_{\ellp h^-}<3.13$\gevcc.  This rejects events where a
muon from a $\jpsi$ decay is misidentified as a pion. The probability
to misidentify a pion as a muon is of the order 2\% and to misidentify
as an electron less than 0.1\%.

We measure the kinematic variables $\mes=\sqrt{s/4 -p^{*2}_B}$ and
$\Delta E = E_B^* - \sqrt{s}/2$, where $p^*_B$ and $E_B^*$ are the \B
momentum and energy in the $\Upsilon(4S)$ CM frame, and $\sqrt{s}$ is
the total CM energy.  For signal events, the \mes\ distribution peaks
at the \B meson mass with a resolution of about 2.5\mevcc, and the
\DeltaE\ distribution peaks near zero with a resolution of about
20\mev, indicating that the candidate system of particles has total
energy consistent with the beam energy in the CM frame. The \B
candidate is required to be in the kinematic region $5.200 <\mes <
5.289 \gevcc$ and $-0.10<\DeltaE<0.05$\gev.

The main backgrounds arise from light quark \qqbar\ continuum events
and \BB\ backgrounds formed from random combinations of leptons from
semileptonic \B and $D$ decays. These are suppressed through the use
of boosted decision tree discriminants (BDTs)~\cite{bib:bdt}. As the
input variable distributions for the \qqbar\ continuum and the \BB\
backgrounds are sufficiently different, two BDTs are trained, one to
distinguish between signal and \qqbar\ continuum, and the other
between signal and \BB\ backgrounds. Each BDT is trained in four
regions according to lepton type (muon versus electron) and mass range
($m_{\ellp\ellp}$ above or below the \jpsi\ mass).  The input
variables consist of \DeltaE\ and seventeen parameters that represent
the event shape of the decay, the distance of closest approach of the
di-lepton system to the beam axis, the vertex probabilities of the
di-lepton and \B candidates, the magnitudes of the thrusts of both the
decay particles and the rest of the event, and the thrust directions
with respect to the beam axis of the experiment.

To construct the BDTs, we use simulated samples of events for the
signal and background, and we assume background decay rates consistent
with measured values~\cite{hfag}. We compare the distributions of the
data and the simulated background variables used as input to the BDTs
and confirm that they are consistent.

The output distributions of the \qqbar\ and \BB\ BDTs are each used to
define probability distribution functions ${\cal P}_{{\rm sig}}$ and
${\cal P}_{{\rm bkg}}$ for signal and background, respectively. The
probabilities are used to define a likelihood ratio \lhratio\ as

\begin{equation}
  \label{eq:lhratio}
\lhratio \equiv \frac{{\cal P}_{{\rm sig}}^{\BB} + {\cal P}_{{\rm sig}}^{\qqbar}}
                  {{\cal P}_{{\rm sig}}^{\BB} + {\cal P}_{{\rm sig}}^{\qqbar} +
       {\cal P}_{{\rm bkg}}^{\BB} + {\cal P}_{{\rm bkg}}^{\qqbar}}\!.
\end{equation}

\noindent We veto candidates if either ${\cal P}_{{\rm sig}}^{\BB}$ or
  ${\cal P}_{{\rm sig}}^{\qqbar}$ is less than 0.5 or the ratio
  \lhratio\ is less than 0.2. This retains 85\% of the simulated signal
  events while rejecting more than 95\% of the background.

After the application of all selection criteria, some events will
contain more than one reconstructed \B candidate. Fewer than 1\% of
accepted events have more than one \B candidate.  We select the most
probable \B candidate from among all the candidates in the event
using the likelihood ratio \lhratio. Averaged over all events, the
correct \B candidate in simulated signal events is selected with greater
than 98.5\% accuracy. For events with more than one \B candidate, the
correct candidate is selected with an accuracy of 67\%-82\%, depending
on the mode. The final event selection efficiency for simulated signal
is 13\%-48\%, depending on the final state.  The selection efficiency
for all modes is approximately constant to within a relative $\pm10\%$
as a function of $m_{\ellp h^-}$ between $m_{h^-}$ and $4.6$\gevcc.

We extract the signal and background yields from the data with an
unbinned maximum likelihood (ML) fit using

\begin{equation}
  \label{eq:ml}
{\mathcal L} = \frac{1}{N!}\exp{\left(-\sum_{j}n_{j}\right)}
\prod_{i=1}^N\left[\sum_{j}n_{j}{\mathcal
    P}_{j}(\vec{x}_i;\vec{\alpha}_j)\right]\!,
\end{equation}

\noindent where the likelihood ${\cal L}$ for each event candidate $i$
is the sum of $n_j {\cal P}_j(\vec x_i; \vec \alpha_j)$ over two
categories $j$: the signal mode $\Bp\to h^-\ellp\ellp$ (including the
small number of misreconstructed \B candidates) and background, as
will be discussed.  For each category $j$, ${\cal P}_j(\vec x_i; \vec
\alpha_j)$ is the product of the probability density functions (PDFs)
evaluated for the $i$-th event's measured variables $\vec x_i$.  The
number of events for category $j$ is denoted by $n_j$ and $N$ is the
total number of events in the sample. The quantities $\vec \alpha_j$
represent the parameters describing the expected distributions of the
measured variables for each category $j$. Each discriminating variable
$\vec x_i$ in the likelihood function is modeled with a PDF, where the
parameters $\vec \alpha_j$ are extracted from MC simulation,
off-resonance data, or on-resonance data with \mes $< 5.27$\gevcc. The
two variables $\vec x_i$ used in the fit are \mes\ and \lhratio.
Since the linear correlations between the two variables are found to
be only 4\%-7\% for simulated signal modes and 8\%-12\% for simulated
background and on-resonance data, we take each ${\cal P}_j$ to be the
product of the PDFs for the separate variables. Any correlations in
the variables are treated later as a systematic uncertainty. The three
free parameters in the fit are the numbers of signal and background
events and the slope of the background \mes\ distribution.

MC simulations show that the \qqbar\ and \BB\ backgrounds have very
similar distributions in \mes\ and \lhratio. We therefore use a single
ARGUS shape~\cite{ArgusShape} to describe the \mes\ combinatorial
background, allowing the shape parameter to float in the fits. The
ratio \lhratio\ for both signal and background is fitted using a
non-parametric kernel estimation KEYS algorithm~\cite{bib:keys}.

\begin{figure}
\begin{center}
\centerline{
  \setlength{\epsfxsize}{1.0\linewidth}\leavevmode\epsfbox{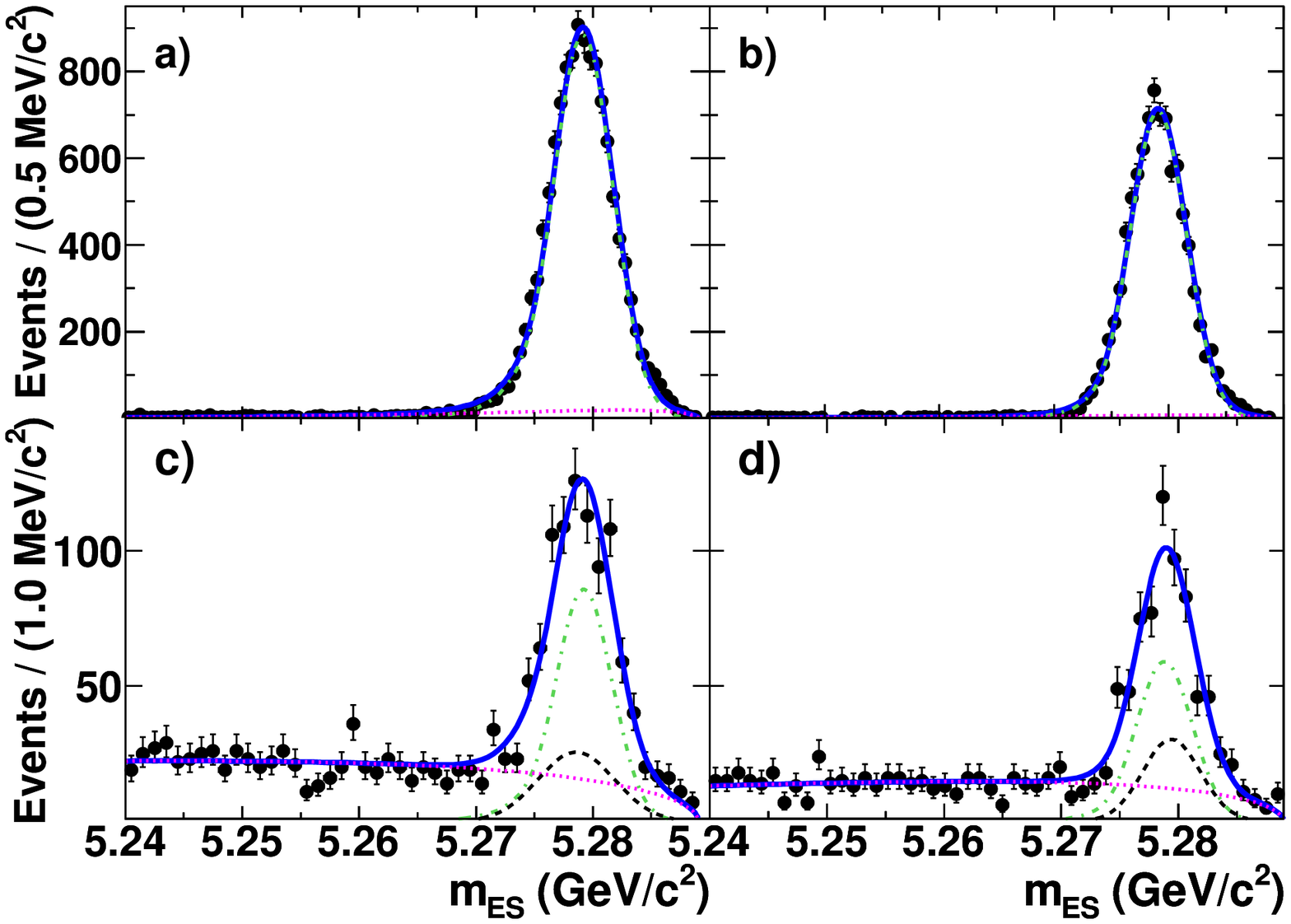}
}
\caption{The \mes\ distributions for 
a) $\Bp\to\jpsi(\to e^+e^-)\Kp$; 
b) $\Bp\to\jpsi(\to \mu^+\mu^-)\Kp$; 
c) $\Bp\to\jpsi(\to e^+e^-)\pip$; 
and d) $\Bp\to\jpsi(\to \mu^+\mu^-)\pip$. 
The solid (blue) line is the
total fit, the dotted (magenta) line is
the background, the dash-dotted (green) line is the signal, the dashed
(black) line is the misidentified $\jpsi\Kp$ events.}
\label{fig:jpsi}
\end{center}
\end{figure}

We parameterize the signal \mes\ distributions using a Gaussian shape
 unique to each final state, with the mean and width determined from
 fits to the analogous final states in the $\Bp\to\jpsi(\to
 \ellp\ellm) h^+$ events from the on-resonance data. The same
 selection criteria as previously given are used, with the modification
 that two opposite-sign leptons are required, the reconstructed
 $\jpsi$ mass must be in the range 2.95 to 3.15\gevcc, \mes\ greater
 than 5.24\gevcc, and \DeltaE between $-0.3$ and 0.2\gev. The signal
 and background \mes\ on-resonance data distributions are fitted with
 a Gaussian and an ARGUS function, respectively. For modes with a pion
 in the final state, we account for $\jpsi\Kp$ misidentified as
 $\jpsi\pip$ by using the signal distribution extracted from the
 $\jpsi\Kp$ data as an additional background. For both $\jpsi\to\epem$
 and $\jpsi\to\mup\mun$, the \jpsi\ mass distribution has a width
 $\sim 15$\mevcc. The \mes\ mean for all modes is
 $5.2791\pm0.0001$\gevcc and the width is $(2.41 - 2.56)$ \mevcc\ with
 an error $(0.02-0.09)$ \mevcc, depending on the mode. The means and
 widths are robust against changes in the assumptions concerning the
 relative contribution of \qqbar\ and \BB\ events to the backgrounds
 and the functions used to fit the signal and background
 distributions. The numbers of measured events for the four modes are
 within one standard deviation of the expected numbers calculated from
 previously measured branching
 fractions~\cite{bib:pdg}. Figure~\ref{fig:jpsi} shows the extracted
 \mes\ distributions for each mode.

As a cross-check of the background PDFs to $\Bp\to h^-\ellp\ellp$, we
perform a fit to a simulated background sample, with the same number
of events as the on-resonance data sample, and also a fit to the
off-resonance data sample. In both cases, the number of signal events
is compatible with zero for all four modes.

We test the performance of the fits to $\Bp\to h^-\ellp\ellp$ by
generating ensembles of MC datasets from both the PDF distributions
and the fully simulated MC events. The mean number of signal and
background events used in the ensembles is taken from the full default
model fit to the selected on-resonance data sample described previously. We
generate and fit 5000 datasets with the number of signal and
background events allowed to fluctuate according to a Poisson 
distribution. The signal yield bias in the ensemble of fits is between
-0.30 and 0.15 events, depending on the mode, and this is subtracted
from the yield taken from the data.

\begin{figure}
\begin{center}
\centerline{
  \setlength{\epsfxsize}{1.0\linewidth}\leavevmode\epsfbox{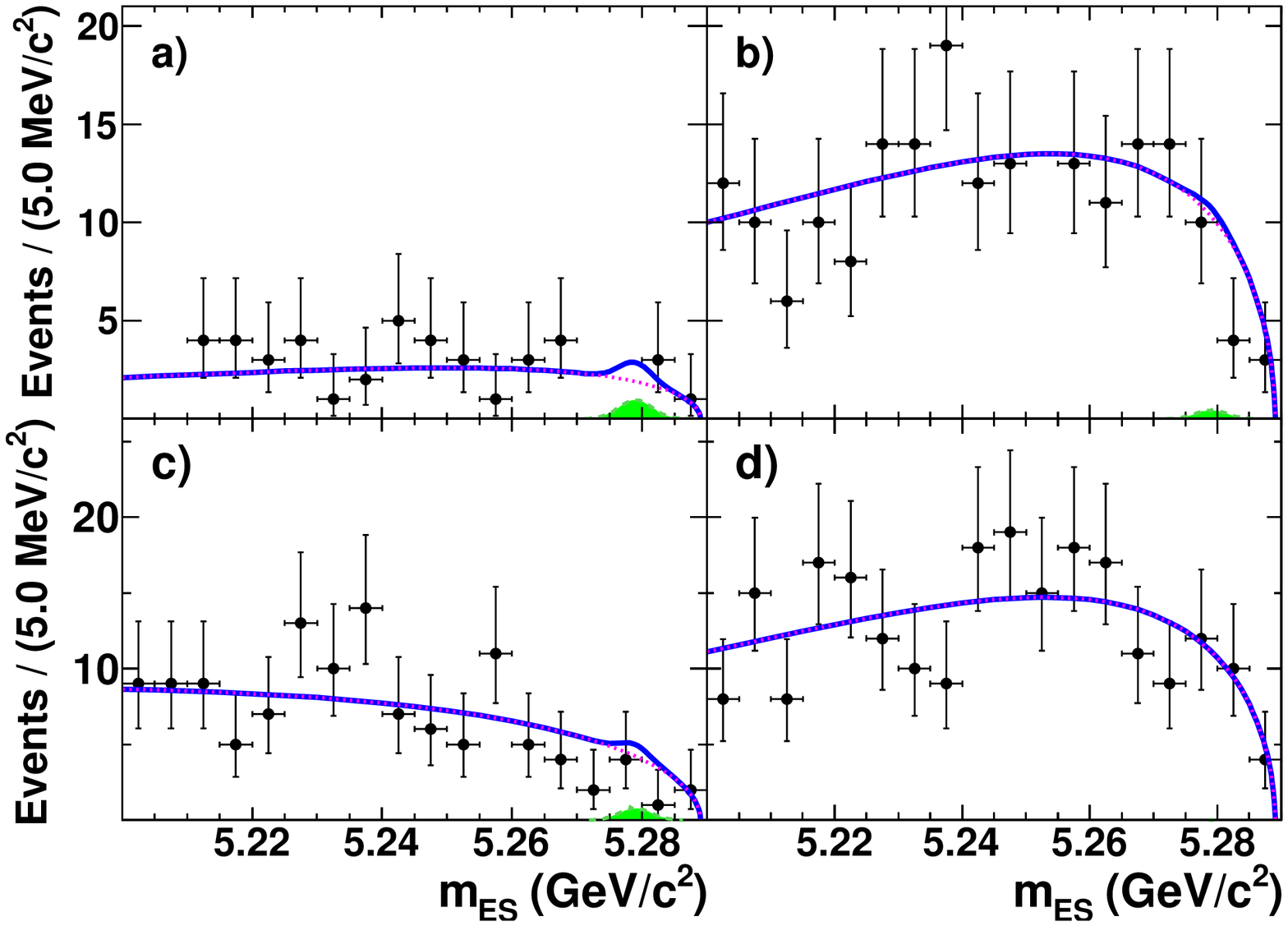}
}
\caption{The \mes\ distributions for 
a) $\BtoKee$; 
b) $\BtoKmm$; 
c) $\BtoPiee$; 
and d) $\BtoPimm$. 
The solid (blue) line is the
total fit, the dotted (magenta) line is
the background, the solid (green) histogram is the signal.}
\label{fig:data}
\end{center}
\end{figure}

The results of the ML fits to the on-resonance data are summarized in
Table~\ref{tab:results}. Figure~\ref{fig:data} shows the \mes\
distributions for the four modes. The signal significance is defined
as $\calS=\sqrt{2\Delta\ln {\cal L}}$, where $\Delta\ln {\cal L}$ is
the change in log-likelihood from the maximum value to the value when
the number of signal events is set to zero. Systematic errors are
included in the log-likelihood distribution by convolving the
likelihood function with a Gaussian distribution with a variance equal
to the total systematic error defined later in this paper. The branching fraction
\BR\ is given by $n_{s}/(\eta N_{\BB})$ where $n_{s}$ is the signal
yield, corrected for the fit bias, $\eta$ is the reconstruction
efficiency and $N_{\BB}$ is the number of \BB\ events collected.

\begin{table*}[htbp!]
\centering
\caption{Results for the measured \B decays, showing the total
  events in the sample, signal yield fit bias (with error), signal
  yield (corrected for fit bias) and its statistical uncertainty,
  reconstruction efficiency $\eta$, significance \calS\ (with statistical
  and systematic uncertainties included), branching fraction \BR,
  and 90\% C.L. branching fraction upper limits $\calB_{UL}$.}
\label{tab:results}
\begin{tabular}{lrrrrrrr}
\hline \hline
Mode & Events & Fit Bias & Yield & $\eta$ (\%) & \calS ($\sigma$) & {\calB}
($\times 10^{-8}$) & $\calB_{UL}$  ($\times 10^{-8}$) \\
\hline
$\Bp\to\pim\ep\ep$   & \nonPiee & \hspace{2mm} \biasPiee &
\hspace{2mm} \nyPieeRound & \hspace{2mm} \effPiee & \hspace{2mm} \sigPieeRound & \hspace{2mm} \bfPiee  & \ulPiee  \\
$\Bp\to\Km\ep\ep$    & \nonKee  & \biasKee &  \nyKeeRound  & \effKee  & \sigKeeRound & \bfKee  & \ulKee  \\
$\Bp\to\pim\mup\mup$ & \nonPimm & \biasPimm &  \nyPimmRound & \effPimm & \sigPimmRound & \bfPimm  & \ulPimm  \\
$\Bp\to\Km\mup\mup$  & \nonKmm  & \biasKmm & \nyKmmRound  & \effKmm  & \sigKmmRound & \bfKmm  & \ulKmm  \\
\hline \hline
\end{tabular}
\end{table*}

The systematic uncertainties in the branching fractions are summarized
in Table~\ref{tab:systematics}.  They arise from the PDF
parameterization, fit biases, background yields, and efficiencies.
The PDF uncertainties are calculated by varying, by their errors, the
PDF parameters that are held fixed in the default fit, taking into
account correlations.  For the non-parametric kernel estimation KEYS
algorithm, we vary the smearing parameter between 50\% and 200\% of
the nominal value.  The uncertainty from the fit bias includes the
statistical uncertainty from the simulated experiments and half of the
correction itself, added in quadrature.

Two tests are used to calculate the contribution to the error caused
by the assumption that the \qqbar\ and \BB\ backgrounds have similar
distributions. We first vary the relative proportions of light quark
\qqbar, \ccbar, and \BB\ used in the simulated background between 0\%
and 100\%. The new simulated background \lhratio\ PDF is then used in
the fit to the data and compared to the default fit to data. We also
perform an ensemble of fits to MC samples consisting of one simulated
signal event and the number of simulated background events given by
the default fit to data. The relative proportions of light quark
\qqbar, \ccbar, and \BB\ in the simulated background are varied and a
fit is performed to the MC sample. The result is compared to the fit to
the default MC sample. The error is calculated as half the difference
between the default fit and the maximum deviation seen in the ensemble
of fits. All the errors described previously are additive in nature and
affect the significance of the branching fraction results.

Multiplicative uncertainties include reconstruction efficiency
uncertainties from tracking (0.8\% per track added linearly for the
leptons and 0.7\% for the kaon or pion), charged lepton particle
identification (0.7\% per track added linearly for electrons, 1.0\%
for muons), hadron particle identification (0.2\% for pions, 0.6\% for kaons),
uncertainty in the BDT response from comparison to charmonium
control samples (2.0\%), the number of \BB\ pairs (0.6\%), and MC
signal statistics (0.2\%). The total multiplicative branching fraction
uncertainty is 3.2\% or less for all modes.

\begin{table}[htbp!]
\begin{center}
\caption[Summary of branching fraction systematics]{Summary of
  branching fraction \BR\ systematic uncertainties for the four decays.}
\label{tab:systematics}
\begin{tabular}{rcccc}
\hline\hline
Systematic           & \Piee & \Kee & \Pimm & \Kmm\\
\hline
\multicolumn{5}{c}{Additive uncertainties (candidates)}\\ 
PDF variation          & $0.01$ & $0.01$ & $0.01$ & $0.09$ \\ 
KEYS PDFs              & $0.30$ & $0.05$ & $0.23$ & $0.02$ \\ 
Fit bias               & $0.09$ & $0.15$ & $0.05$ & $0.04$ \\
Backgrounds            & $0.05$ & $0.07$ & $0.25$ & $0.35$ \\
Total                  & \addPiee & \addKee & \addPimm & \addKmm \\ \hline
\multicolumn{5}{c}{Multiplicative uncertainties (\%)}\\
Lepton tracking        & $1.6$  & $1.6$  & $1.6$ & $1.6$ \\
Hadron tracking        & $1.4$  & $1.4$  & $1.4$ & $1.4$ \\
Lepton ID             & $0.7$  & $0.7$  & $1.0$ & $1.0$ \\
Hadron ID             & $0.2$  & $0.6$  & $0.2$ & $0.6$ \\
BDT                    & $2.0$  & $2.0$  & $2.0$ & $2.0$ \\
\BB\ pairs             & $0.6$  & $0.6$  & $0.6$ & $0.6$ \\
MC statistics          & $0.2$  & $0.2$  & $0.2$ & $0.2$ \\ 
Total                 & $3.1$& $3.1$  & $3.2$ & $3.2$ \\ \hline
\multicolumn{5}{c}{Branching fraction \BR\ uncertainties $(\times 10^{-8})$}\\
Additive              & $0.14$ & $0.12$  &  $0.56$ & $0.34$ \\
Multiplicative        & $0.01$ & $0.02$  &  $0.01$ & $0.01$ \\
Total                 & $0.14$ & $0.12$  &  $0.56$ & $0.35$\\ 
\hline\hline
\end{tabular}
\end{center}
\end{table}

As shown in Table~\ref{tab:results}, we observe no significant yields.
The 90\% C.L. branching fraction upper limits $\calB_{UL}$ are
determined by integrating the total likelihood distribution (taking
into account statistical and systematic errors) as a function of the
branching fraction from 0 to $\calB_{UL}$, such that
$\int^{\calB_{UL}}_{0} \calL d\calB = 0.9\times\int^{\infty}_{0} \calL
d\calB$. The upper limits are dominated by the statistical error.

\begin{figure}
\begin{center}
\begin{tabular}{c}
   \epsfig{file=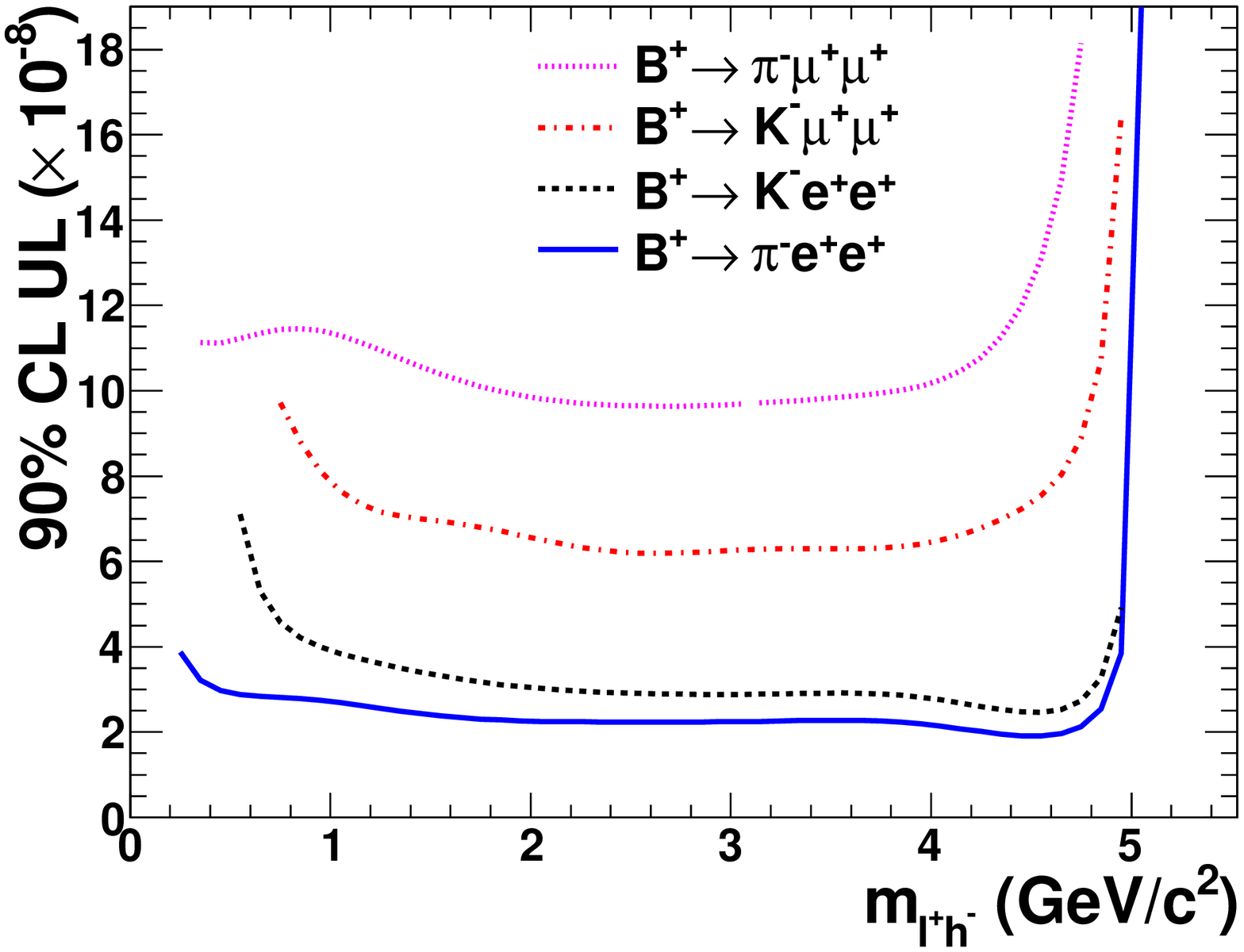,width=1.0\columnwidth} 
\end{tabular}
\caption{90\% C.L. upper limits on the branching fraction as a
 function of the mass $m_{\ellp h^-}$ for \BtoPimm\ (dotted/magenta
 line), \BtoKmm\ (dash-dotted/red line), \BtoKee\ (dashed/black line),
 and \BtoPiee\ (solid/blue line).}
\label{fig:effic}
\end{center}
\end{figure}

Figure~\ref{fig:effic} shows $\calB_{UL}$ as a function of the mass
$m_{\ellp h^-}$ for the four modes. The $\calB_{UL}$ limit is
recalculated in bins of 0.1\gevcc with the assumption that all the
fitted signal events are contained in that bin. The total likelihood
distribution from the default fit is rescaled taking into account the
reconstruction efficiency in each $m_{\ellp h^-}$ bin and the
increased uncertainty in the estimate of the reconstruction efficiency
due to reduced MC statistics. The $\calB_{UL}$ limit in each $m_{\ellp
h^-}$ bin is then recalculated using the formula given above.  The
change in shape is mainly due to the variation of the reconstruction
efficiency as a function of the mass. If the decay $\Bp\to h^-
\ellp\ellp$ is caused by the exchange of a Majorana neutrino, as
illustrated in Fig.~\ref{fig:feyn}, then $m_{\ellp h^-}$ can be
related to the Majorana neutrino mass $m_{\nu}$~\cite{bib:atre}.

In summary, we have searched for the four lepton-number violating
processes $\Bp\to h^-\ellp\ellp$. We find no significant yields and
place 90\% C.L. upper limits on the branching fractions in the range
$(2.3-10.7)\times 10^{-8}$. The branching fraction upper limit for
\BtoPimm\ is less restrictive than the result reported in
Ref.~\cite{bib:lhcb}, while the \BtoKmm\ limit is commensurate. The
limits for \BtoKee\ and \BtoPiee\ are 30 and 70 times more stringent,
respectively, than previous measurements at \epem colliders~\cite{bib:cleo}.

\input acknowledgements.tex


\end{document}

%% file: mysymbols.tex
\def\lhratio {{\ensuremath{\cal R}}\xspace}

\newcommand{\offreslumi} {\mbox{43.9\invfb}}
\newcommand{\nbb}        {\mbox{$471\pm3$}}

\def\Bmeson  {\B\ meson}

\def\Piee    {\ensuremath{\pim\ep\ep}}
\def\BtoPiee {\ensuremath{\Bp\to\Piee}}
\def\effPiee {\ensuremath{47.8\pm0.1}}
\def\biasPiee {\ensuremath{+0.15\pm0.09}}

\def\nonPiee {\ensuremath{123}}

\def\nyPieeRound {\ensuremath{0.6^{+2.5}_{-2.7}}} 
\def\sigPieeRound {\ensuremath{0.4}} 
\def\addPiee {\ensuremath{0.32}} 
\def\bfPiee {\ensuremath{0.27^{+1.1}_{-1.2}\pm0.1}} 
\def\ulPiee {\ensuremath{2.3}} 

\def\Kee    {\ensuremath{\Km\ep\ep}}
\def\BtoKee {\ensuremath{\Bp\to\Kee}}
\def\effKee {\ensuremath{30.9\pm0.1}}
\def\biasKee {\ensuremath{-0.30\pm0.15}}

\def\nonKee {\ensuremath{42}}

\def\nyKeeRound {\ensuremath{0.7^{+1.8}_{-1.2}}}
\def\sigKeeRound {\ensuremath{0.5}} 
\def\addKee {\ensuremath{0.17}} 
\def\bfKee {\ensuremath{0.49^{+1.3}_{-0.8}\pm0.1}} 
\def\ulKee {\ensuremath{3.0}}  

\def\Pimm    {\ensuremath{\pim\mup\mup}}
\def\BtoPimm {\ensuremath{\Bp\to\Pimm}}
\def\effPimm {\ensuremath{13.1\pm0.1}} 
\def\biasPimm {\ensuremath{-0.01\pm0.05}}

\def\nonPimm {\ensuremath{228}}

\def\nyPimmRound {\ensuremath{0.0^{+3.2}_{-2.0}}}
\def\sigPimmRound {\ensuremath{0.0}} 
\def\addPimm {\ensuremath{0.34}} 
\def\bfPimm {\ensuremath{0.03^{+5.1}_{-3.2}\pm0.6}} 
\def\ulPimm {\ensuremath{10.7}}  

\def\Kmm    {\ensuremath{\Km\mup\mup}}
\def\BtoKmm {\ensuremath{\Bp\to\Kmm}}
\def\effKmm {\ensuremath{23.0\pm0.1}}
\def\biasKmm {\ensuremath{+0.02\pm0.04}}

\def\nonKmm {\ensuremath{209}}

\def\nyKmmRound {\ensuremath{0.5^{+3.5}_{-2.5}}}
\def\sigKmmRound {\ensuremath{0.2}} 
\def\addKmm {\ensuremath{0.35}} 
\def\bfKmm {\ensuremath{0.45^{+3.2}_{-2.7}\pm0.4}} 
\def\ulKmm {\ensuremath{6.7}}  



%% file: authors_dec2011.tex
%
\author{J.~P.~Lees}
\author{V.~Poireau}
\author{V.~Tisserand}
\affiliation{Laboratoire d'Annecy-le-Vieux de Physique des Particules (LAPP), Universit\'e de Savoie, CNRS/IN2P3,  F-74941 Annecy-Le-Vieux, France}
\author{J.~Garra~Tico}
\author{E.~Grauges}
\affiliation{Universitat de Barcelona, Facultat de Fisica, Departament ECM, E-08028 Barcelona, Spain }
\author{A.~Palano$^{ab}$ }
\affiliation{INFN Sezione di Bari$^{a}$; Dipartimento di Fisica, Universit\`a di Bari$^{b}$, I-70126 Bari, Italy }
\author{G.~Eigen}
\author{B.~Stugu}
\affiliation{University of Bergen, Institute of Physics, N-5007 Bergen, Norway }
\author{D.~N.~Brown}
\author{L.~T.~Kerth}
\author{Yu.~G.~Kolomensky}
\author{G.~Lynch}
\affiliation{Lawrence Berkeley National Laboratory and University of California, Berkeley, California 94720, USA }
% ...
\collaboration{The \babar\ Collaboration}
\noaffiliation

%% file: acknowledgements.tex
We are grateful for the excellent luminosity and machine conditions
provided by our \pep2\ colleagues, 
and for the substantial dedicated effort from
the computing organizations that support \babar.
The collaborating institutions wish to thank 
SLAC for its support and kind hospitality. 
This work is supported by
DOE
and NSF (USA),
NSERC (Canada),
CEA and
CNRS-IN2P3
(France),
BMBF and DFG
(Germany),
INFN (Italy),
FOM (The Netherlands),
NFR (Norway),
MES (Russia),
MICIIN (Spain),
and STFC (United Kingdom). 
Individuals have received support from the
Marie Curie EIF (European Union)
and the A.~P.~Sloan Foundation (USA).